# Polymer-based Capacitive Micromachined Transducer-Enabled Inline Monitoring of Ultrasonic Welding in Thermoplastic Carbon Fiber Composites

Jonas Welsch & Dominik Görick, Martin Angerer, Jinhao Lu, Sergei Vostrikov, Michael Kupke, Heinz Voggenreiter, Andrea Cossettini, Luca Benini, Edmond Cretu, Robert Rohling

J. Welsch, M. Angerer, J. Lu, E. Cretu, R. Rohling
Electrical & Computer Engineering, The University of British Columbia, Vancouver, Canada
E-Mail: j.welsch@ece.ubc.ca

D. Görick, M. Kupke, H. Voggenreiter
Center for Lightweight-Production-Technology, Deutsches Zentrum für Luft und Raumfahrt e.V., Augsburg, Germany

S. Vostrikov, A. Cossettini, L. Benini
Integrated Systems Laboratory, ETH Zürich, Zürich, Switzerland

**Funding:** This work was sponsored in part by the Natural Sciences and Engineering Research Council of Canada (RGPIN-2020-05061) and in part by the Innovate BC Ignite Program (IGNITE-2024-RND14-1378-UBC-Rohling-Sonus) as well as the Federal Ministry for Research, Technology and Aeronautics (01IS22018B) of the Federal Republic of Germany.

Welsch and Görick contributed equally to this work and should be considered shared first authors with Welsch being named first and acting as corresponding author.

**Abstract:** Thermoplastic composite structures enable lightweight, recyclable, and high-throughput aerospace manufacturing, but reliable quality assurance of advanced joining processes remains a key challenge. This work presents a compact, low-cost, and wireless ultrasonic non-destructive testing system for real-time, inline monitoring of continuous ultrasonic welding of thermoplastic carbon fiber composites. The system integrates custom-fabricated polymer-based capacitive micromachined ultrasonic transducers (polyCMUTs) with the ultra-low-power WULPUS platform, enabling operation in the harsh, high-interference welding environment. An eight-element linear polyCMUT array operating at a center frequency of approximately 3.6 MHz is designed, fabricated, packaged, and integrated into an industrial welding setup. Inline measurements are performed during welding of carbon fiber laminates with intentionally introduced defects. Process-synchronous ultrasonic data reveal consistent depth-of-echo shifts at defect locations, in strong agreement with X-ray computed tomography ground truth. Across 21 welds, all induced defects are detected without false negatives and with limited false positives. The results demonstrate that polymer-based CMUT technology enables robust, scalable, and manufacturing-compatible ultrasonic sensing, representing a decisive step toward intelligent process monitoring and quality assurance for next-generation thermoplastic composite welding.

## 1. Introduction

The aerospace industry is projected to experience unprecedented growth, with Airbus and Boeing each forecasting the delivery of over 43,000 new aircraft within the next two decades. [1,2] To limit environmental impact and operating costs, reducing fuel consumption is paramount. The primary solution, other than increased engine efficiency and improved aerodynamics, is a significant reduction in structural weight. [3,4]

Recent aircraft models are incorporating an ever-increasing percentage of light weight carbon fiber reinforced plastics (CFRPs) as their primary structural material, replacing traditional high-performance aluminum alloys. [5]

Among CFRPs, thermoplastic matrix materials are attracting growing interest due to their advantages, including improved chemical resistance, impact tolerance, recyclability and up to 50 % possible weight reduction. Advances in the manufacturing of thermoplastic composites (TCs) also offer significantly shorter fabrication times, with reductions of up to 80 %, further lowering production costs. [4,6,7]

Composite manufacturing techniques can also reduce the number of joints in a structure, but they cannot eliminate them entirely. [8] Traditional mechanical joining methods such as bolts and rivets not only add several tons to a finished aircraft, but also pose challenges for quality control, rely on complex manual processes, and require through-holes that can compromise the composite by disrupting the fiber layup. [8-12]

The ability of TCs to melt and resolidify enables more advanced joining techniques, such as continuous ultrasonic welding (CUW). This method eliminates the need for additional fasteners like metal rivets and allows for rapid, automated, and in-line compatible bonding. [13,14]

CUW is well-suited for aerospace applications due to its potential for high throughput, reduced cost, and compatibility with TCs. It operates by applying high-frequency mechanical vibrations through a sonotrode, generating localized heating via interfacial friction and viscoelastic damping. This melts the thermoplastic matrix and forms a strong bond under pressure. [15,16] However, due to the complexity and novelty of the process, industry adoption has been slow [13], and ensuring the reliability of CUW joints remains a significant challenge. Therefore, non-destructive testing (NDT) methods capable of real-time, inline weld quality assessment are essential to enable large-scale industrial deployment. [17-19]

Ultrasonic testing (UT) is well-established in aerospace for detecting subsurface defects such as porosity and delamination and has been investigated for thermal welding of thermoplastic materials before. [20] However, conventional piezoelectric transducers are often bulky, expensive, and ill-suited for integration into automated and harsh manufacturing environments like the continuous ultrasonic welding scenarios, due to their fragility, rigid form factor [21] and their tendency of being affected by lower frequency harmonic resonances outside the intended center frequency. [19] Capacitive micromachined ultrasonic transducers (CMUTs) offer a promising alternative. CMUTs are essentially microscopic capacitive membranes, fabricated by micromachining silicon wafers. Initially developed for medical imaging, CMUTs feature a wide bandwidth, improved acoustic coupling, and miniaturization via lithography-based cleanroom fabrication. [22] During CMUT operation a DC-Bias is applied between a fixed bottom electrode and a top electrode within or on-top of a movable membrane. The created electrostatic forces deflect the membrane to reduce the gap and enhance sensitivity. By applying an AC pulse tailored to the resonance frequency of the membrane, the change in electrostatic force leads to membrane vibration and a transmitted ultrasonic acoustic signal. In receive, the acoustic signal will deflect the biased membrane, creating a measurable change in capacitance and a flow of charges. The CMUT form factor, ranging from sub millimeter to a few centimeters, and silicon wafer-based fabrication allow for seamless integration into compact systems with embedded electronics. [23]
As an alternative to standard silicon-based CMUTs, polymer-based CMUTs (polyCMUTs) have further advanced this technology by simplifying fabrication, reducing costs, and reducing fabrication time, therefore enabling rapid prototyping approaches to system design. [19,24,25] This paper presents the development of a compact, wireless, and cost-effective NDT system based on polyCMUTs and the Wearable Ultra-Low Power Ultrasound (WULPUS) open-source platform developed at ETH Zürich. [26] The completed system is custom designed to evaluate weld quality in thermoplastic composite joints formed by CUW in real-time during the ongoing process, and represents a step toward intelligent, adaptive manufacturing in aerospace. The work includes in-house polyCMUT fabrication, custom electronics development, and experimental validation in both laboratory and industrial CUW environments.

# 2. Methods

## 2.1. System Overview

The following sections describe the complete development of a compact ultrasonic NDT system, from transducer design and fabrication to system integration and deployment in an industrial CUW environment. Transducer development and production were performed entirely at the University of British Columbia (UBC), while system-level testing took place at the Center for Lightweight Production Technology (ZLP) of the German Aerospace Center (DLR) in Augsburg, Germany.
The fabrication process consisted of three key stages: first, the microfabrication of polymer-based CMUT arrays; second, post-processing and packaging into functional transducer assemblies; and third, integration with custom electronics for real-time ultrasonic inspection. These stages are detailed below.

## 2.2. Transducer Array Design

The polyCMUT transducer array was tailored for seamless integration into the CUW process and full compatibility with the WULPUS system. WULPUS is an ultra-low-power, 8-channel (time-multiplexed) ultrasound probe that supports a sampling rate of 8 MSps per channel, ±30 V bipolar excitation, and a 30 V DC bias, necessitating a transducer design with high sensitivity at limited driving voltages and maximum 8 MHz operating frequency. [26]
A linear array with eight elements was selected and designed to leverage all available channels while providing signal strength along the weld seam. Each element measures 1.1 mm in width and 8.3 mm in length, with a 520 µm pitch. Within each element, 540 circular membrane cells (96 µm diameter, 8 µm inter-cell spacing) are arranged in six parallel columns.
The array's center frequency was targeted between 3 and 4 MHz. Though this frequency is close to the Nyquist limit of the maximum sample rate, it optimizes weld-zone resolution and reduces the spectral overlap between the welding and inspection frequency. Resonance and acoustic behavior was modeled using a hybrid analytical-numerical approach, [27] incorporating cell geometry, membrane thickness, and surrounding medium properties. [22,28] A 5 µm laminate thickness was selected, yielding a predicted collapse voltage of 57 V, tailored to WULPUS constraints. This represents the voltage at which the membrane is pulled into contact with the bottom electrode as the electrostatic force overcomes the restoring force.[29] All design characteristics were chosen as a balance between reliable and high-yield fabrication and the required imaging performance. A summary of the array's design parameters is provided in **Table 1**.

| Parameter | Value |
|---|---|
| *Cell diameter* | 96 µm |
| *Cell-to-cell pitch* | 8 µm |
| *Electrode coverage* | 85 % |
| *Element size* | 8.3 x 1.1mm |
| *Cell count per element* | 540 |
| *Element pitch* | 520 µm |
| *Laminate thickness* | 5 µm |
| *Collapse voltage* | 57 V |
| *Center frequency* | 3.8 MHz |

**Table 1.** Chosen design parameters for the custom 8 channel polyCMUT sensor.

### 2.3. Microfabrication

The polyCMUT arrays were fabricated using a low-temperature, additive process adapted from established recipes, with optimizations for low-voltage operation and simplified processing. [28] An outline can be found in **Figure 1**. All design masks were created in L-Edit (Tanner EDA, MEMS 2021).

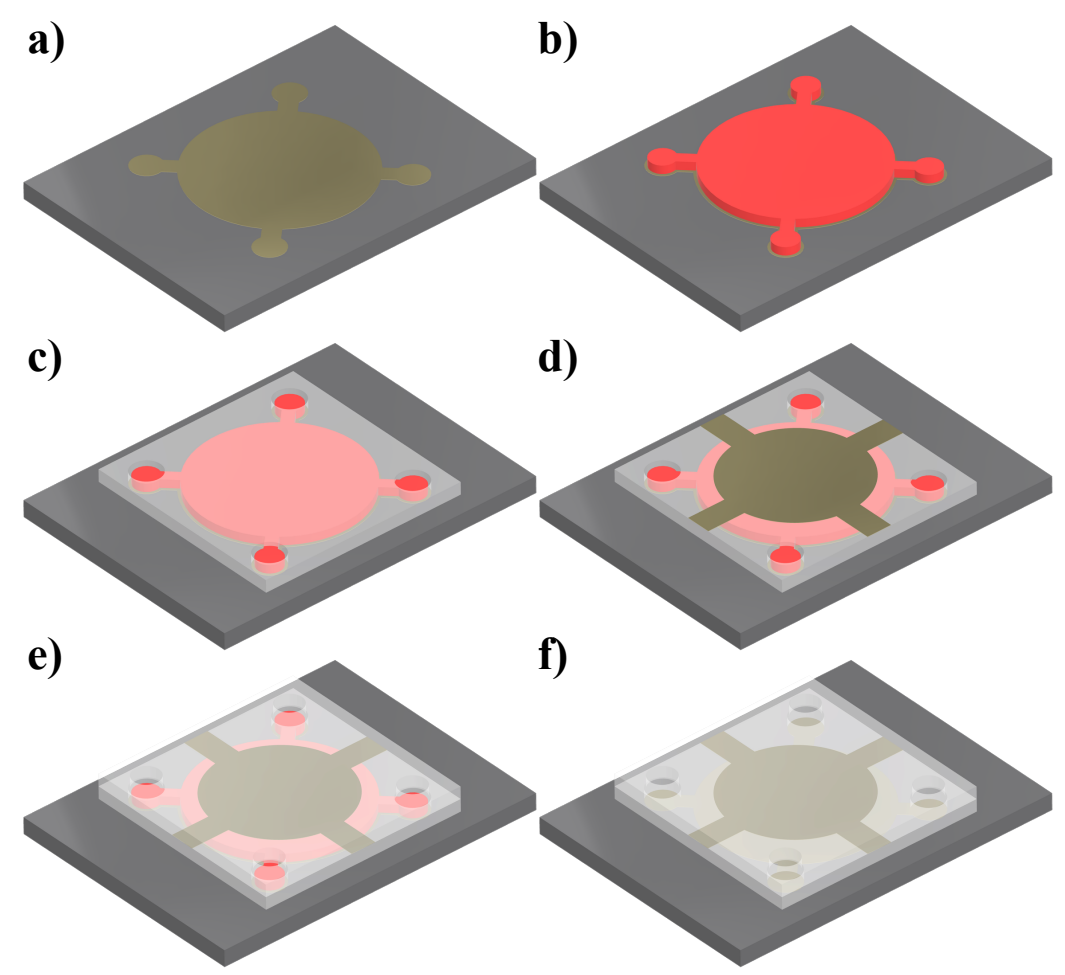

**Figure 1.** Cross-section of polyCMUT manufacturing steps on the example of a single cell. a) patterning of gold bottom electrode, b) patterning of LOR 1A sacrificial layer, c) patterning of SU8 2000.5 membrane layer, d) deposition of gold top electrode, e) patterning of top membrane layer, f) creation of gap via wet etch of sacrificial layer.

Fabrication began on a 4-inch, 500 µm thick silicon wafer coated with 500 nm thermal $SiO_2$ as an insulating substrate. Bottom electrodes were patterned using nLOF 2020 (Merck KGaA, Germany) photoresist and exposed via maskless lithography (MLA 150, Heidelberg Instruments) at 350 nm, 110 mJ/cm².

After development and hard bake, a 5 nm Ti, 90 nm Au, 5 nm Ti layer was deposited via electron beam evaporation (AJA UHV Hybrid Evaporator). Lift-off was completed in acetone followed by IPA and DI water cleaning.
A bilayer sacrificial layer was then created using LOR 1A (Kayaku Advanced Materials, USA) and nLOF 2020, resulting in a layer thickness after lift-off of 170 nm (measured with a Dektak XT Profilometer). Patterning was performed with identical lithography settings, and the masking layer was removed in acetone.
The first SU-8 membrane layer (SU-8 2000.5, Kayaku Advanced Materials) was spin-coated at 4000 rpm, yielding ~440 nm thickness. It was exposed, developed, and hard-baked according to manufacturer protocols. The top electrode (5 nm Ti / 90 nm Au / 5 nm Ti) was then deposited. A second SU-8 membrane layer (SU-8 2005, Kayaku Advanced Materials, ~4.2 µm) was applied and patterned. All wafers were diced (Disco DAD3204) and immersed in AZ 300 MIF to release the membranes by dissolving the sacrificial layer. Final drying was performed in a critical point dryer (Autosamdri-815B, Tousimis) to avoid stiction.
The entire process including etch and release was conducted within a 4-day period. **Table 2** provides an overview of all layer thicknesses used in the final stack and **Figure 2** shows the finished transducer.

| Layer | Material | Thickness |
|---|---|---|
| *Bottom electrode* | Ti / Au / Ti | 5 / 90 / 5 nm |
| *Sacrificial layer* | LOR 1A | 170 nm |
| *Membrane Layer I* | SU8 2000.5 | 440 nm |
| *Top Electrode* | Ti / Au / Ti | 5 / 90 / 5 nm |
| *Membrane Layer II* | SU8 2005 | 4200 nm |

**Table 2.** Layer overview and thicknesses of the fabricated polyCMUTs. All thickness values were measured multiple times across the wafers and averaged.

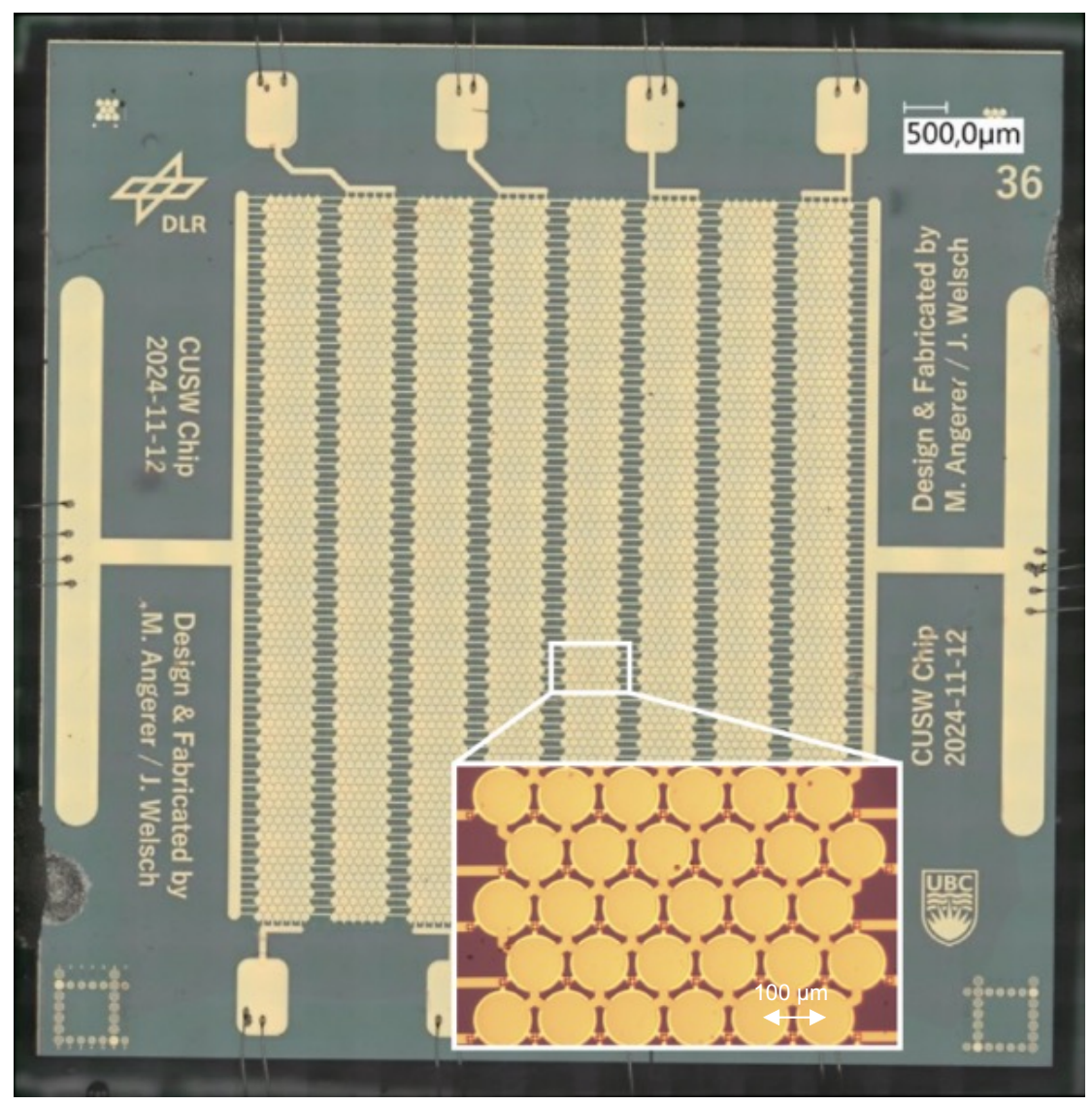

**Figure 2.** Image of polyCMUT chip with 5x magnification of a single element.

### 2.4. Post-processing

After dicing and sacrificial release, the polyCMUT dies underwent post-processing steps to be reliably used in-situ as shown in **Figure 3**.
The chips were bonded onto a custom-designed 4-layer PCB using a low-outgassing epoxy (EpoTek 301-2, Epoxy Technology Inc.), with the layout matching the footprint and pitch of the 8-channel array. Electrical interfacing was established via 25 µm aluminum wire bonds. A UV-curable adhesive (9001-E-V3.5, Dymax Inc) was applied to secure and protect the bond wires.
To protect the exposed CMUT structures and etch holes, the assembly was coated with 4 µm of Parylene-C. A polydimethylsiloxane (PDMS, Sylgard 184, Dow Chemical Company) encapsulation layer was added to further protect the membranes and create a smooth contact for acoustic coupling. The PDMS was cast using a thin, 3D-printed mold to ensure uniform thickness and to limit signal attenuation. **Figure 4** shows the resulting system, the polyCMUT array and a shielded coaxial ribbon connector.

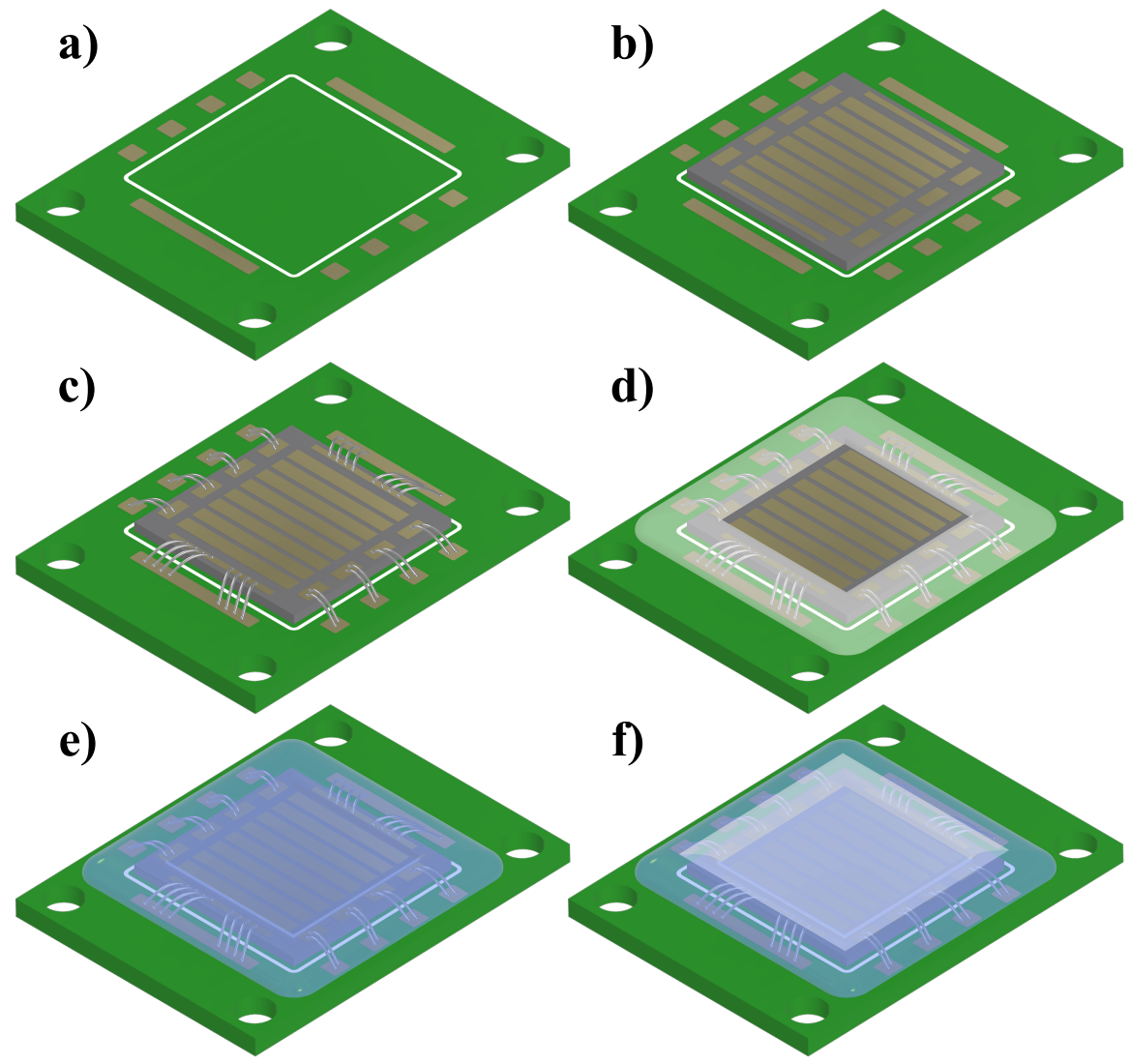

**Figure 3.** Postprocessing steps: a) diced polyCMUT die, b) chip is bonded onto 4 layer PCB, c) wire bonding to connect contact pads, d) encapsulation of bonds with partial glob top, e) Parylene-Coating to seal membranes, f) PDMS protective layer.

### 2.5. WULPUS Integration

The finalized polyCMUT frontends were integrated with the WULPUS system. To accommodate the need for stable DC biasing, a custom shield was developed featuring a passive bias-tee network composed of a high-impedance resistor and a DC-blocking capacitor. This ensured a consistent, on-board generated membrane bias of 30 V while isolating front-end electronics. Compared to the original design, the excitation voltage was doubled (from 15 V to 30 V). [26]
The shield was implemented as a compact 4-layer PCB that stacked directly onto the WULPUS mainboard as seen in Figure 4. Shielded coaxial flex-cables (Samtec, USA) were used to interface the sensor PCB and WULPUS, which reduces electromagnetic interference and enables placement of the WULPUS away from the harsh welding environment. During a single measurement, 7 channels were transmitting while one separate channel operated in receiving mode to maximize transmit power and minimize potential multiplexer switching artifacts.

The complete system, including the sensor, biasing shield, and WULPUS, was compact enough to be embedded within the limited space of the welding fixture. The WULPUS main PCB and shield, including the cable connector, measured 25 × 47 × 16 mm, while the polyCMUT PCB, also including its cable connector, measured just 20 × 43 × 10 mm. The wireless capability of WULPUS (which transmits raw data via BLE) obviates the need for data cables and enables testing in areas which are otherwise difficult to reach.

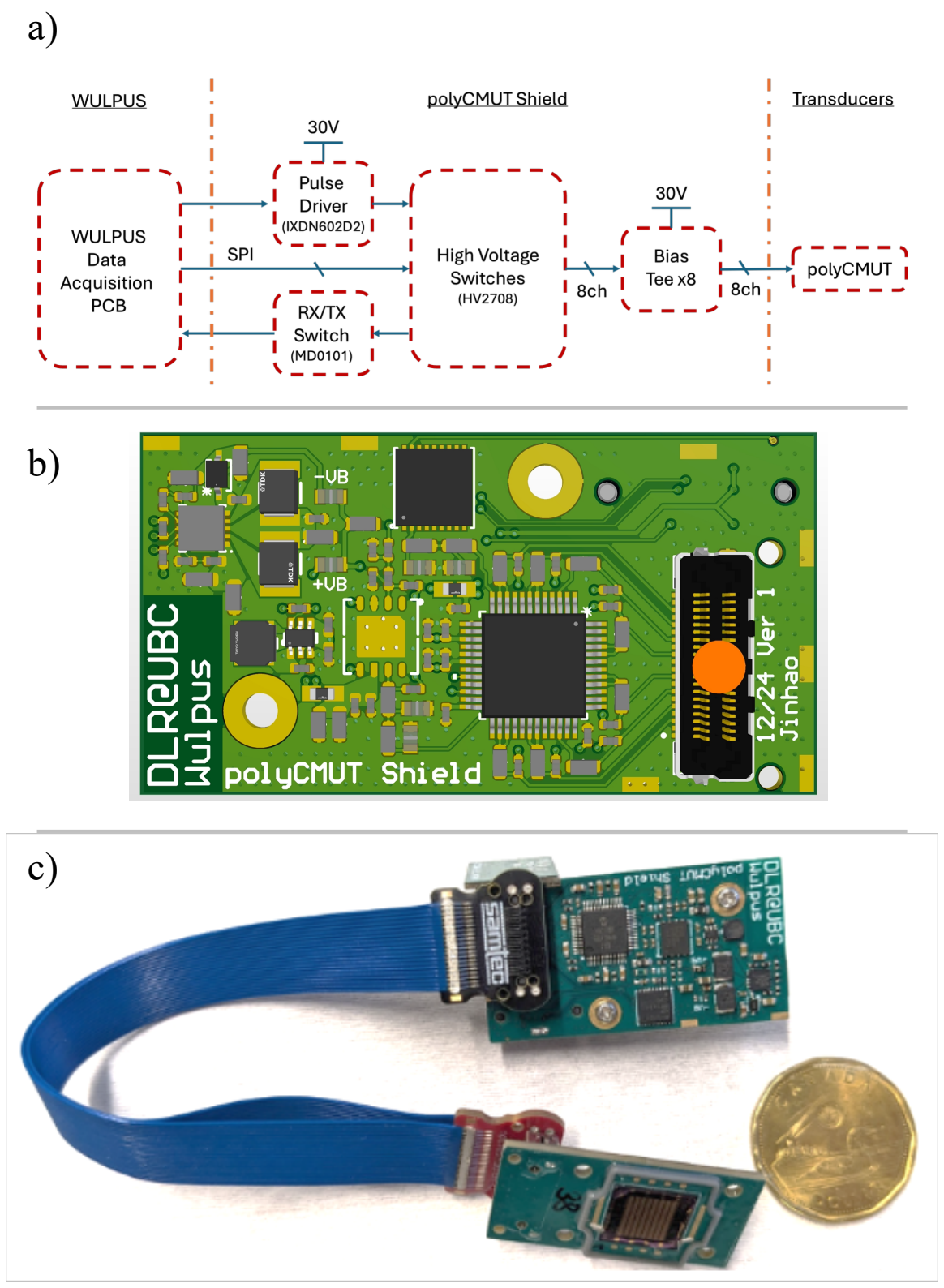

**Figure 4.** a) WULPUS polyCMUT shield system overview, b) Screenshot of polyCMUT shield PCB in Altium Designer, c) WULPUS system with a custom polyCMUT shield and the fabricated 8 channel sensor attached via coaxial cable - Canadian one dollar coin as reference.

### 2.6. Ultrasonic Welding Setup

Thermoplastic composite samples from T700G carbon fiber laminates with a $(0/90)_{3s}$ layup and a Low-Melting Polyaryletherketone (LMPAEK) matrix were acquired and cut to size to create 21 welding samples. A narrow strip of matrix material, exceeding the width of the weld seam, was attached to one of the plates above the weld seam using a soldering iron. This strip will hereafter be referred to as the energy director since it allows for a greater amount of matrix material in the welding zone and improves weld quality. [30] The plates measured 296 mm in length and 1.68 mm in thickness and were joined with a 12.7 mm overlap using CUW. The setup is depicted in **Figure 5**.

A compaction roll applied 300 N of contact force, while the sonotrode delivered 800 N during welding. A rectangular heat sink, immediately following the sonotrode, compresses the laminate with 16 bar pressure and cools the welding zone. This allows the TC matrix within the weld to resolidify under continuous pressure, resulting in a thin weld seam. Welding power was set to 1050–1100 W at a velocity of 15 mm/s and 20 kHz welding frequency. The polyCMUT frontend was mounted to the rear of the heat sink via two steel bars. The sensor was acoustically coupled to the weld seam through a 12 mm Polyetheretherketone (PEEK) block, which matched the acoustic impedance of the laminate. [19] The block not only protects the sensor surface from abrasion through the rough CFRP surface but also creates an acoustic spacer, allowing for more time for the acoustic field to form and for WULPUS to switch between send and receive mode.

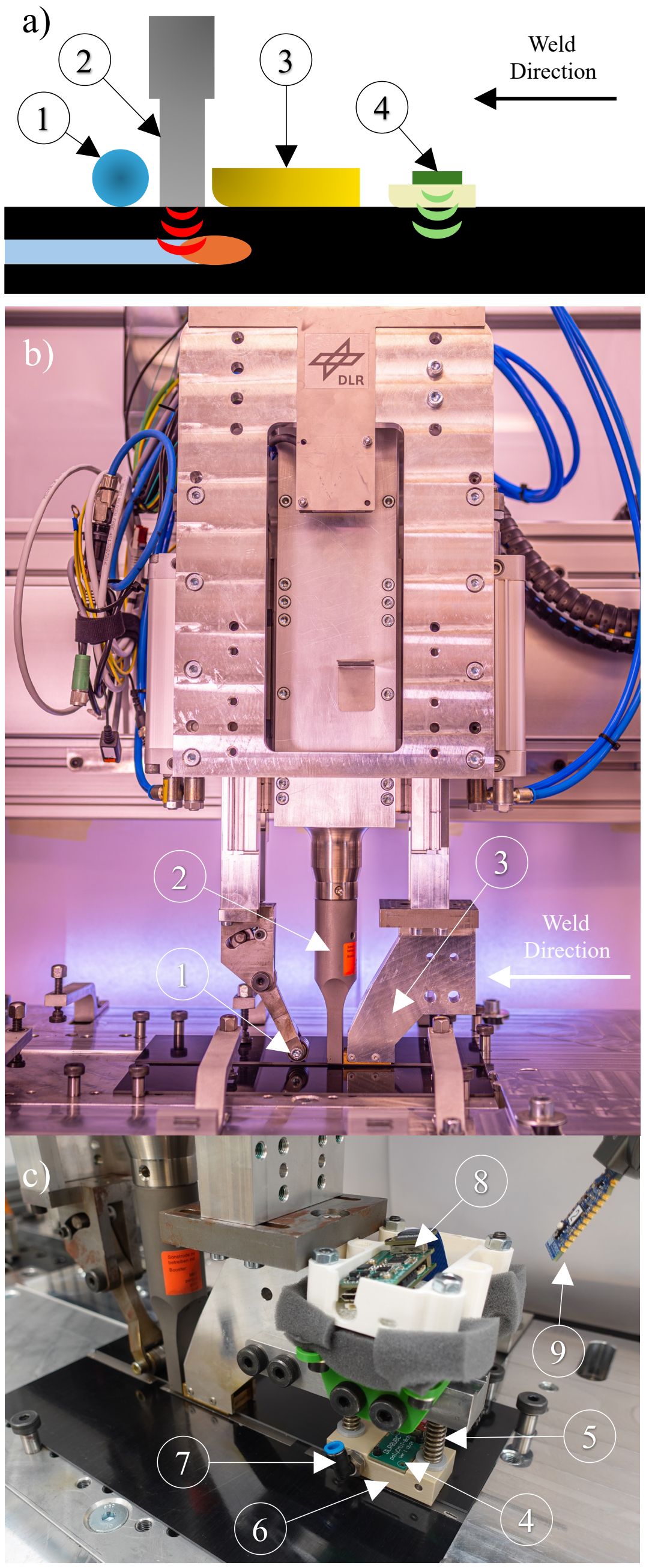

**Figure 5.** a) Schematic of the Ultrasonic welding process with cross-section of the weld seam visualizing the melting energy director and joining of the TC. b) Welding System without polyCMUT system, c) Close-up view of the WULPUS and polyCMUT mounting on the welding system. 1) compaction roll, 2) sonotrode, 3) heat sink, 4) polyCMUT transducer, 5) PEEK spacer block, 6) suspension spring, 7) glycerin hose connector, 8) WULPUS & polyCMUT shield, 9) WULPUS Bluetooth dongle.

Two springs (VD-207KK, Gutekunst Federn, Germany) applied ~16.6 N to the PEEK block, creating a suspension system that guarantees continues contact while being able to accommodate potential debris or surface roughness. Glycerin was delivered into milled out channels on the underside of the PEEK block via a peristaltic pump, ensuring a constant thin film of glycerin, for ultrasound coupling and acoustic impedance matching. [19]
Measurement parameters were set using the WULPUS user interface with the ADC start delay configured to acquire data from 28mm (before the weld seam) to 380mm, with a sampling interval of 0.02s. With a scanning range of 352 mm and the chosen sampling interval, each weld produced ~1173 measurements, corresponding to a spatial resolution of ~300 µm. The system was synchronized with the TwinCAT-controlled welding system via ADS and Python scripting.

**2.7. Defect Localization**

To evaluate sensitivity to defects, weld errors were introduced by omitting sections of the energy director, which prevented bonding in those areas (**Figure 6**). In the first weld, defects were located at 100–110 mm and 200–210 mm. A second sample (as well as all samples used in subsequent trials) confirmed reproducibility, with defects at 106.9 mm and 216.5 mm (middle of the error, error width 10 mm).
Sample 1 was analyzed using X-ray computer tomography (CT) (v|tome|x L240/450, Waygate Technologies, Germany) to visualize internal weld quality and establish a ground truth of defect size and location for comparison with ultrasound data. 20 additional samples with defects at the same positions were prepared for all further tests.

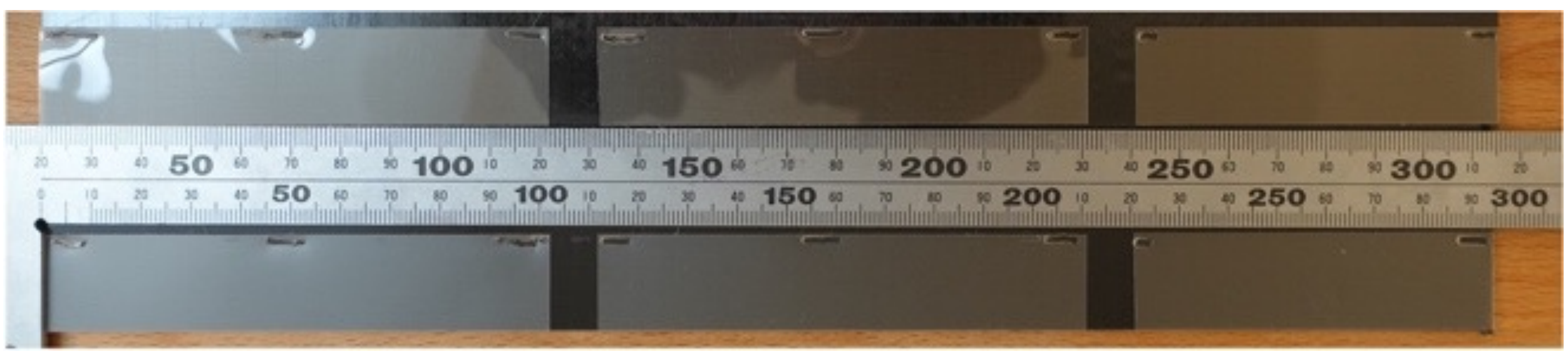

**Figure 6.** CFRP plate with cut energy director to create artificial welding defects as used in the 20 sample welds.

# 3. Results and Discussion

This section presents the results obtained with the developed polyCMUT-based ultrasonic NDT system. First, we assess the overall transducer quality and fabrication consistency using electrical and acoustic performance metrics. These quality control measures provide insight into yield, uniformity, and expected behavior of the fabricated transducer. Next, we report results from initial integration tests conducted within the ultrasonic welding setup, without activating the welding process, to evaluate coupling quality, signal integrity, and transducer sensitivity under ideal conditions. Finally, we present real-time measurements acquired during continuous ultrasonic welding of thermoplastic composite parts and discuss the repeatability and reliability of the system.

### 3.1. Quality Control

Electrical impedance spectroscopy was performed on newly fabricated polyCMUT arrays to assess device functionality, fabrication yield, and electro-mechanical uniformity. Measurements were conducted after membrane release using a network impedance analyzer (HP 4294A, Keysight, USA) and a manual needle prober. An open-short-load calibration procedure was implemented following the method in Teston et al. to minimize parasitic effects from the measurement setup.[31] Two key metrics were extracted: the low-frequency input capacitance $C_p$ at 0.75 MHz and the membrane resonance frequency $f_R$ at a DC bias of 20 V. **Figure 7** displays the impedance spectrum of a single transducer as well as boxplots of the two metrics from 14 arrays (112 total elements). Of these, four elements exhibited short circuits (impedance < 100 Ω), resulting in a fabrication yield of 96%. In terms of variability, $f_R$ shows a standard deviation below 1% of the mean across all measured elements. However, the boxplots indicate that the standard deviation within each transducer array is lower than the overall variation, on average by a factor of 3.6. These results highlight the robustness and repeatability of the fabrication process. While $f_R$ was largely consistent across all devices, one transducer array (#1) showed elevated $C_p$ values with higher variations. Since all arrays share the same design, these results suggest that higher fabrication tolerances affected array #1. Notably, $f_R$ was unaffected, indicating that $C_p$ may serve as a more sensitive parameter for quality control in future implementations.

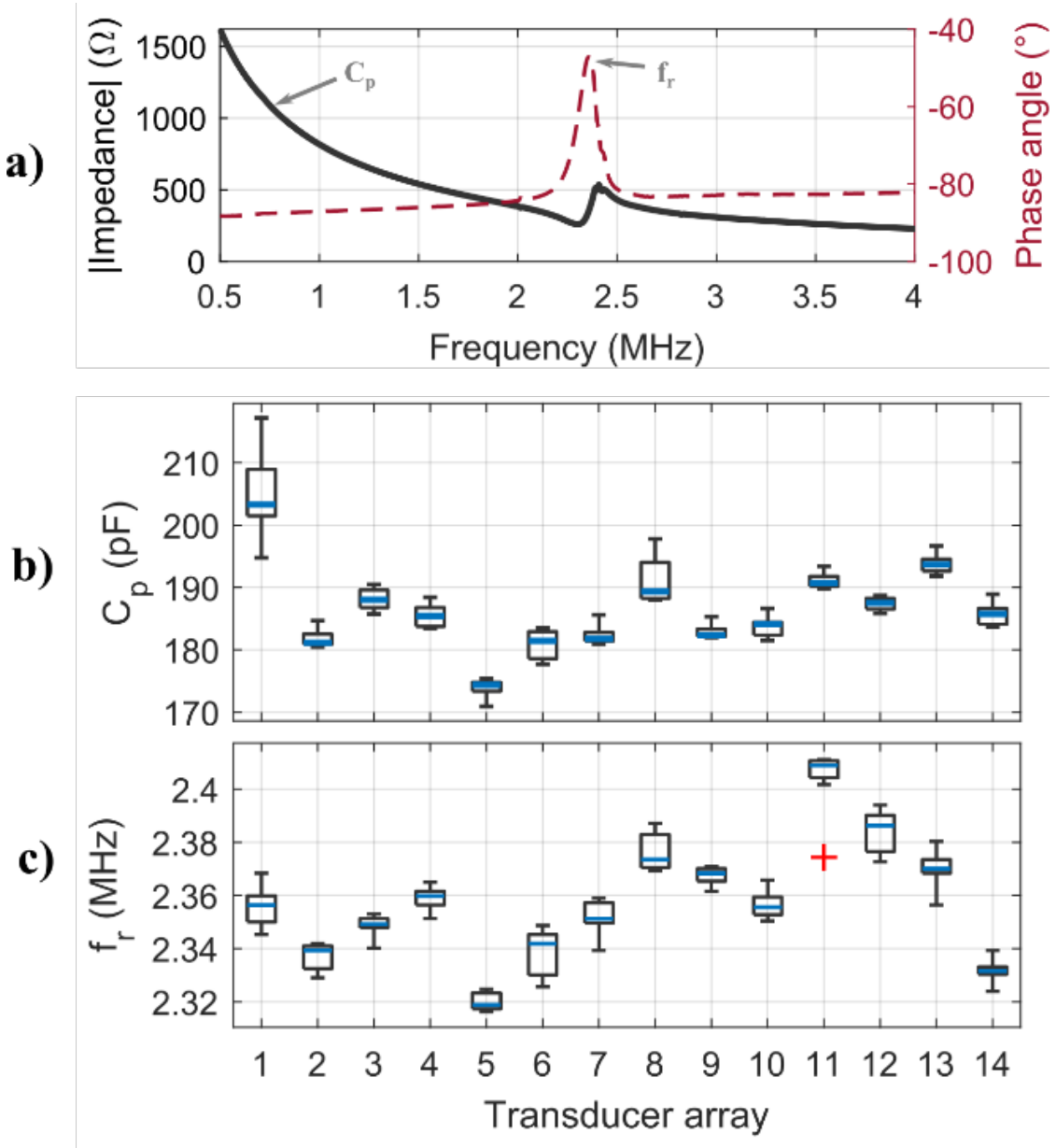

**Figure 7.** Results from electrical impedance spectroscopy: (a) exemplary spectrum; (b) and (c), Boxplots of input capacitance and resonance frequency from 14 transducer arrays (8 channels each).

### 3.2. Performance Characteristics

The acoustic performance of the fabricated polyCMUT arrays was evaluated via underwater transmission measurements. Several arrays were mounted to test PCBs and submerged in a water tank. A calibrated needle hydrophone (HNC-1000, Onda, USA) on a motorized XYZ-stage recorded the radiated acoustic far-field at a radial distance of 10 cm. Excitation was performed using a 20 µs Gauss-weighted linear chirp (0.4–10 MHz, 20 V peak-to-peak), with a DC bias of 40 V. Recorded signals were corrected for hydrophone sensitivity and pulse shape to obtain absolute acoustic sensitivity (Pa/V).
To characterize the angular response, the acoustic transmit pressure was measured over a ±45° arc. **Figure 8** shows the measurement setup, the transmit sensitivity across 16 elements, and an example of the frequency- and angle-dependent response. The latter allows to assess the generated acoustic beam in 2D. Center frequency $f_C$, fractional bandwidth (FBW), and peak sensitivity $S_C$ are listed in **Table 3**.

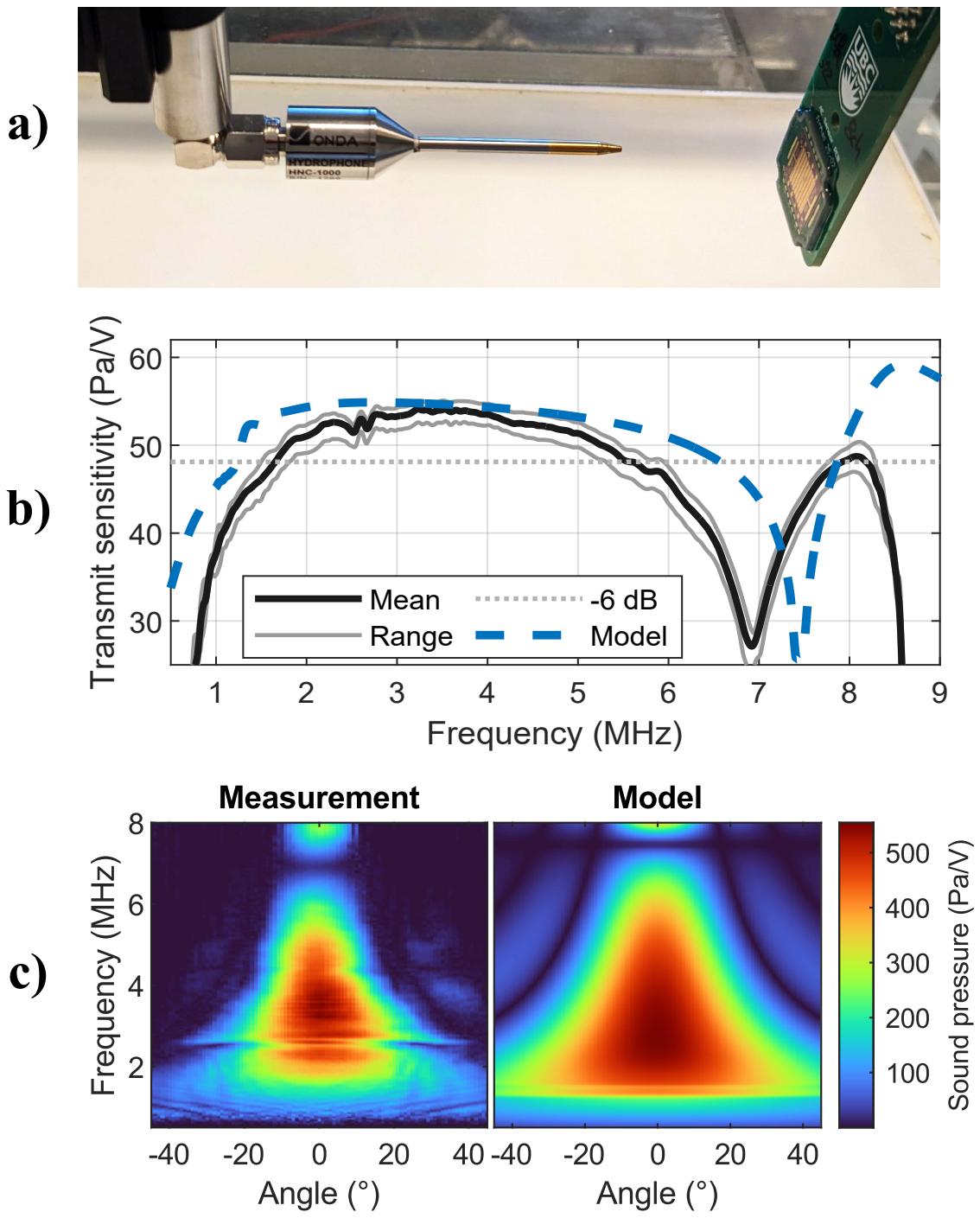

**Figure 8.** Transducer performance characteristics at 10 cm axial distance: (a) Measurement setup; (b) Frequency response from 16 elements compared with model predictions; (b) comparison of frequency- and angle-dependent response between exemplary measurement and model predictions.

The transducers demonstrated a broadband response with a measured mean FBW of 109%. The characteristic notch near 7 MHz stems from the complex interaction between the membrane vibration mode and the acoustic load impedance, leading effectively to an almost zero output. [27] The notch was observed at slightly lower frequencies than predicted, arising from different mechanical boundary conditions than assumed. The measured angular response was narrower than the modeled pattern, suggesting the effective aperture is larger than the physical aperture.
Collapse voltage measurements confirmed safe operating margins: the observed collapse voltage was 65 V, slightly exceeding the target of 57 V and safely above the 30 V bias applied in operation. This result affirms the devices' suitability for integration with the WULPUS system.

| Parameter | Model | Measurement |
|---|---|---|
| *$f_C$ [MHz]* | 3.8 | 3.62 ± 0.08 |
| *FBW [%]* | 135 | 109 ± 2.5 |
| *$S_C$ [Pa/V]* | 554 | 515 ± 44 |
| *$V_{pullin}$ [V]* | 57 | ~ 65 |

**Table 3.** Figures of merit (mean and standard deviation) of the fabricated polyCMUT arrays.

### 3.3. Pretesting & System Evaluation

Prior to *in-situ* testing, acoustic characterization of the TC plates was performed using a three-step protocol. This not only allows for the overall evaluation of system performance but also allows to obtain the speed of sound of the TC, a value that needs to be determined experimentally due to the inhomogeneous nature of composites.[32] The polyCMUT sensor was mounted on a PEEK block and used to collect signals from three configurations:

- Backwall echo of the standalone PEEK block in air coupled top the sensor with ultrasound gel.
- PEEK block acoustically coupled to one composite plate with glycerin.
- PEEK plus two composite plates coupled with glycerin to simulate a complete lap joint.

For each configuration, several pulse-echo measurements were acquired. All signals exhibited the same characteristics. For each case the location of the maximum amplitude was measured and taken as the main reflection of the feature in question. (highlighted in red and overlayed to a typical signal in **Figure 9**).

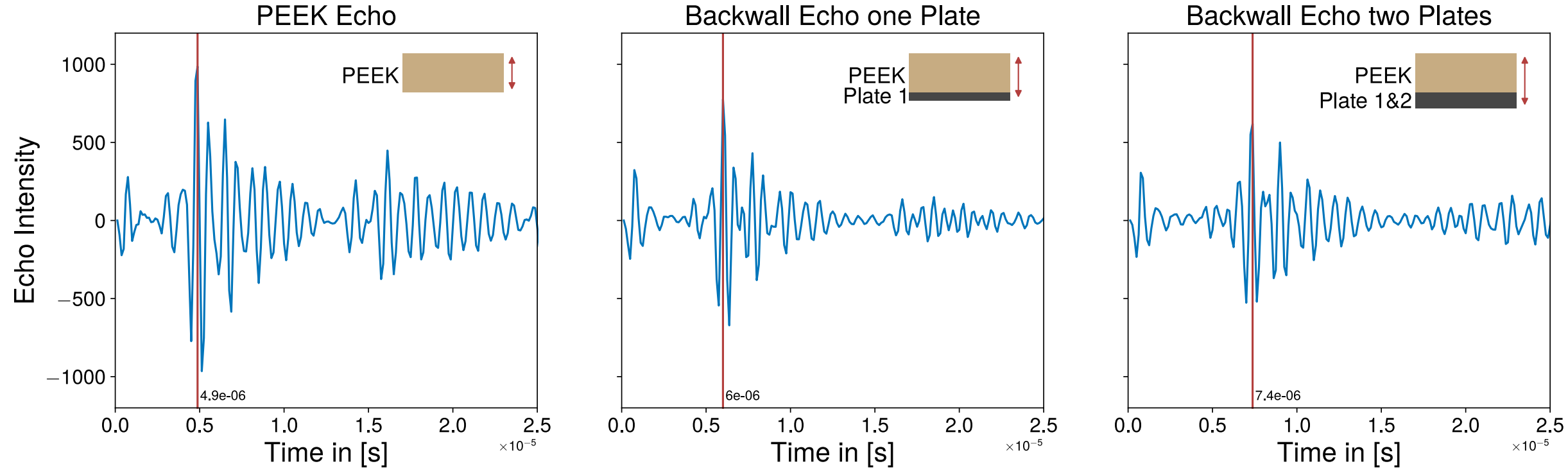

**Figure 9.** Evaluation of signal time of flight depending on material to validate welding seam location. a) signal with PEEK coupling block, b) signal with PEEK and first CFRP plate, c) signal of PEEK block and both coupled CFRP plates.

A single plate shifted the echo by 1.125 µs at a sampling rate of 8 MSps, corresponding to a speed of sound of 2986 m/s at a plate thickness of 1.68 mm. The second plate had an average shift of 1.3 µs with the extra time of flight being attributed to the glycerin matching layer representing the welding seam.

The calculated speed of sound matches values typical for high stiffness TCs and the results allow for the definition of an echo depth specific to the weld zone.[32]

The system, including the PEEK spacer, was next used on pre-welded specimen with introduced defects and consistently showed a repeating pattern: Areas of good weld quality would produce the signal of maximum amplitude at 8 µs, corresponding to a reflection from the backside of the lower TC plate according to the previous trials. Areas of introduced defects would show their strongest reflections at 6.3 µs, corresponding to the location of the welding zone, or the backside of the top plate. This effect was then utilized to plot the images of **Figure 10** and to serve as a simple distinction between good and bad welding quality.

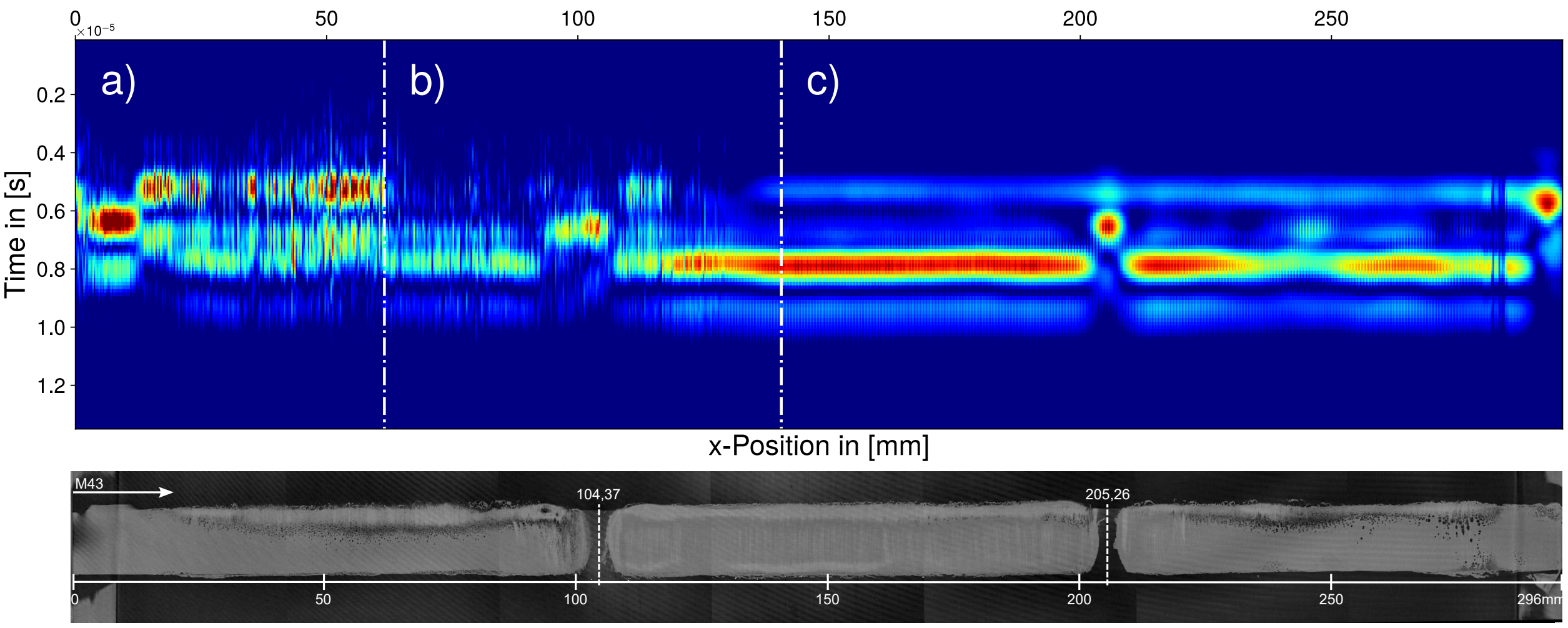

**Figure 10.** Top: Welding signals plotted over position with color coded intensity over depth of echo (similar to a B-Scan). Section a) is not usable due to unreliable surface contact, Section b) is a measurement during normal welding operation and section c) is the measurement while the sonotrode was already shut off. Bottom: High resolution CT scan of the same specimen with porosities visible after the second defect.

### 3.4. Real-Time Monitoring During Welding

Real-time A-scans recorded during welding contained mechanical and electro-magnetic interference from the active welding system. To suppress these components and enhance echo detection, a digital bandpass filter of 30$^{th}$ order (1–3.6 MHz) included in the WULPUS interface, was applied. The absolute value of the Hilbert transform is calculated to get the shape of the measured signals. To isolate echoes from depth, a Gaussian filter centered at µ = 5.875 µs and a standard deviation = 2.5 µs was applied (see Equation 1). For visualizations like Figure 13 the filter is multiplied by 0.8 in order to get a smoother color distribution.

$$G = \frac{12}{\sigma\sqrt{2\pi}} e^{-\frac{1}{2}\left(\frac{x-\mu}{\sigma}\right)^2} \quad (1)$$

The processed output of each A-scan was then visualized in a spatially ordered color plot resembling a continuous B-scan (a 2D cross-sectional image composed of all individual measurements), where red indicates strong echo signals and blue represents weak or absent reflections (Figure 10).

Comparison with CT scans of the welded sample reveals strong agreement. The CT cross-section (greyscale) shows material presence in grey and voids in black. Defects at 104.37 mm and 205.26 mm are clearly visible, while regions before the first and after the second defect show porosity. Both defects show an average width of ~ 5.4 mm. These features correlate with amplitude variations and time shifts in the ultrasound data. Specifically, at 100 mm and 205 mm, signal shifts and depth changes are consistently observed, matching with the expected error locations after accounting for sound propagation through the layered laminate. The welding test stands range of welds is limited in length to reduce cost and infrastructure requirements of larger sample size trials. The mechanical design of the ultrasound inspection sled attached to the welding rig then leads to a creation of 3 specific zones of data as visible in Figure 10.

The first section of the scan (Section (a), up to ~ 70 mm) creates unusable data due to loss of surface coupling of the PEEK spacer to the plates as visible by the considerable amounts of maximum amplitudes at a depth of 5 μs, correlating with the results in Figure 9. In the very initial stage of this section the PEEK spacer is still moving onto the TC and due to the welding zone not starting immediately, the plates begin to vibrate. This vibration is then transferred to the ultrasound system, leading to undesired full system vibration and consecutively, to a deteriorated signal.

In the second part of the scan (Section (b) in **Figure 10**, up to ~140 mm), signals still show considerable electrical and mechanical interference and some coupling failures but reliably produce results even in the noisy environment. Section (b) is therefore of main interest for this work. The relatively short length of 70 mm was considered sufficient due to the number of samples welded overall.

After 140 mm, due to the distance of the ultrasound testing assembly to the sonotrode, the welding is shut off. The measurement was continued to the end of the plate, and the recorded data (Section (c)) can be included to establish a baseline functionality of the system, though will not be considered for detailed evaluation.

### 3.5. Reproducibility of Measurement

To evaluate measurement reproducibility, 20 weld trials were performed using different polyCMUT transducers under identical conditions. Rather than visualizing all results as B-scans we extracted the time of maximum amplitude from each A-scan (individual signal plots) and plotted the resulting graph for each weld. A sliding average filter with a window size of 15 samples was applied to reduce noise.

**Figure 11** shows the overlay of all times of maximal amplitudes plotted over the measurement location. As mentioned in the previous paragraphs, Sections (a) and (c) will be disregarded. **Figure 12** shows a close-up view of section (b). Despite the present interference, each profile exhibits distinct depth of echo changes at the defect locations. These deviations consistently align with the introduced defects, though a small positional variation remains due to propagation effects in the multi-layer laminate and varying thicknesses of the ultrasonic coupling gel between the polyCMUT and the PEEK block.

Crucially, the signal shape consistently changes at the defect position of roughly 100 mm across all trials, confirming that both the sensor arrays and the measurement protocol provide robust, repeatable detection of weld defects.

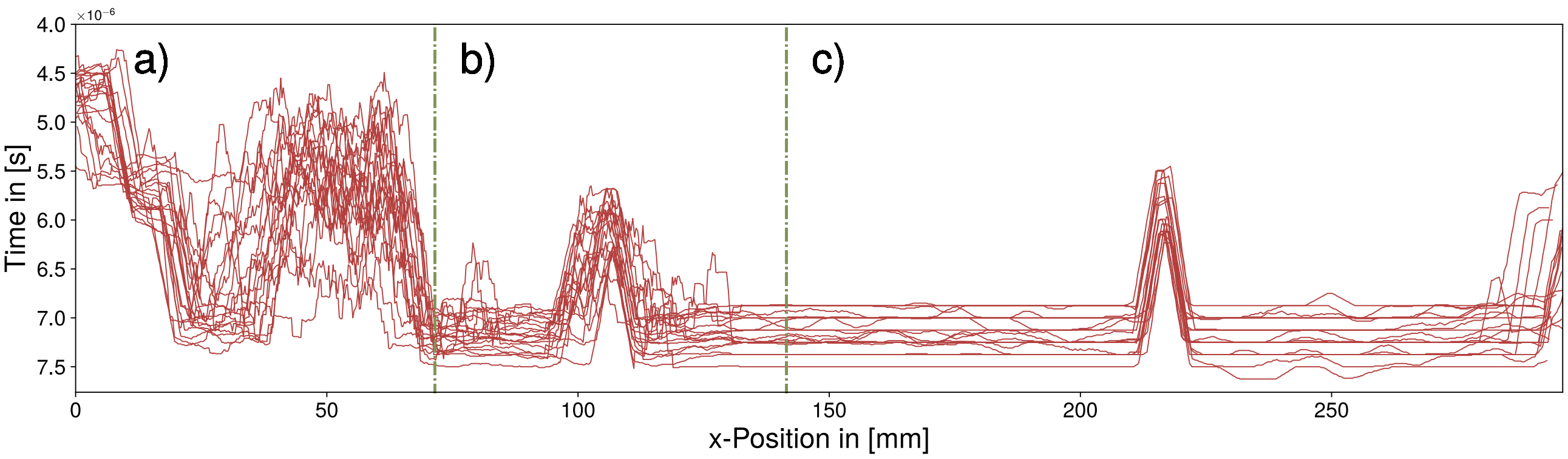

**Figure 11.** A-scan maxima of 20 welds plotted over positions on the plates and corresponding depth with a sliding average filter of 15 for noise reduction. In section a) the NDT assembly is moving onto the welding specimen and does not see reliable contact, in section b) the welding process is running, and the ultrasonic testing is operating as intended and in section c) the NDT system is measuring the remainder of the specimen without the sonotrode operating.

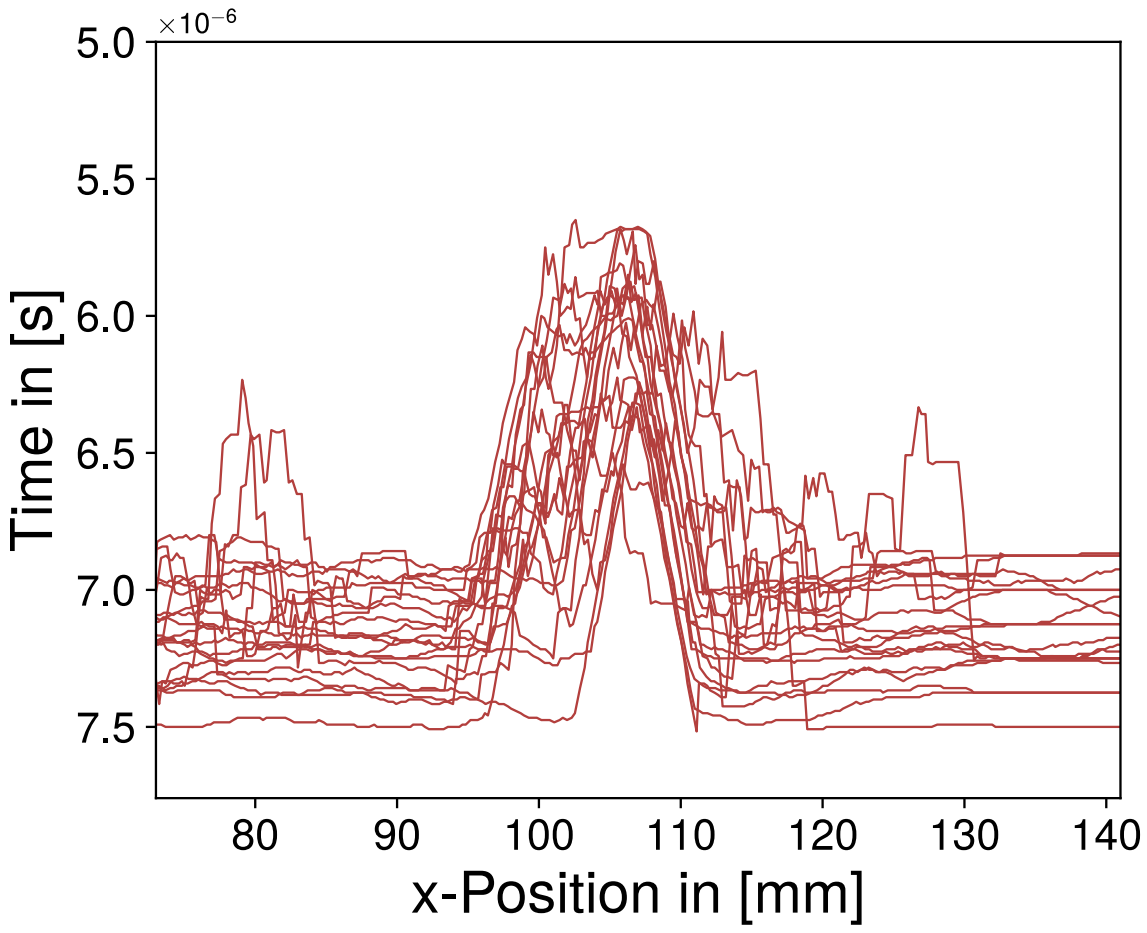

**Figure 12.** Section (b) - measurement during the running welding process with the four false positives clearly visible at 75 and 115 mm. All 20 measurements captured the intentionally induced defect.

### 3.6. System Evaluation & Future Work

In the evaluation of non-destructive testing (NDT) systems, two types of misidentifications are commonly considered: false negatives, where the system fails to detect an existing defect, and false positives, where the system indicates a defect that is not present. As described previously, the system did not miss any defects, thus producing no false negatives. Regarding false positives, Figure 12 shows three peaks at 72 mm and 120 mm, respectively, that could be considered false positives. Two of these peaks exhibit echo depths below 6.5 µs, indicating that they originate from within the lower plate rather than the welding zone. These are disregarded, as plate defects themselves are outside the scope of this investigation. A result with no false negatives and only two potential false positives falls within the acceptable parameters for a reliable NDT system, as outlined in the commonly used standard MIL-HDBK-1823A. [33]

In addition to its overall performance, the system is low-cost as well as compact enough to be fully integrated within the welding setup close to the sonotrode. Moreover, it was shown to withstand the harsh conditions of ultrasonic welding. Transducer design, fabrication, integration, and testing were fully carried out at UBC within four months. Testing and integration into the welding environment were completed at ZLP within two months, resulting in a total development and fabrication cycle of under six months and a material cost of less than 500 USD. The open architecture and fast adaptability of WULPUS and the polyCMUT process enable rapid future system improvements and customization.

Current limitations of the system and conducted tests include the short scan length during welding, lack of variability in defect size and type, and residual signal interference. Producing a larger set of samples with varied defect characteristics would enable proper statistical evaluation of the NDT system and the development of probability of detection (POD) assessments for different defect sizes. [33] Enhanced shielding of the transducers and WULPUS, along with improved acoustic design and signal filtering could further increase signal strength and allow for more precise defect detection.

Future work will also include the investigation of plate defects and weld seam porosity. Porosity significantly compromises structural integrity, and the initial measurement in **Figure 10**, between 230 mm and 250 mm, demonstrates promising potential for its detection using the polyCMUT system.

# 4. Conclusion

In this work, we presented a compact, custom-designed wireless ultrasound system for inline monitoring of ultrasonic welding in thermoplastic composites. The system integrates a low-cost sensor frontend (polyCMUTs) with a configurable backend platform (WULPUS), providing an ideal tool for quality control and monitoring of novel joining processes. In-situ testing showed promise in finding defects, with no false-negative detections for the tested samples (n=21). The polyCMUTs were able to reliably transmit and receive ultrasonic signals in close proximity to the sonotrode despite the harsh and high-interference environment. Hence, it shows promise as a reliable and fully automated in-line NDT system which could lead to more widespread industry adoption of the ultrasonic welding process of thermoplastic composites and hence, be a decisive enabling factor for next-generation lightweight aircrafts.

**Acknowledgment**

The authors would like to thank the UBC Advanced Materials and Process Engineering Laboratory (AMPEL) Nanofab, the UBC Bioimaging Facility and the UBC Centre for Flexible Electronics and Textiles (CFET) for supporting the polyCMUT fabrication. We would also like to thank CMC Microsystems for providing needed software access, as well as Raouf Jemmali and the DLR Institute of Structures and Design in Stuttgart for taking the CT images.

**Conflict of Interest**

The authors declare the following financial interests/personal relationships which may be considered as potential competing interests: Edmond Cretu, and Robert Rohling are the Founder and the Director of Sonus Microsystems, and Robert Rohling is the Founder and an Executive of Sonic Incytes.

**Declaration of use of AI**

The authors declare that artificial intelligence tools were used solely for language editing, text refinement, and formatting support during the preparation of this manuscript. No AI tools were used to generate scientific content, analyze data, interpret results, or create figures. All scientific decisions, data analyses, and conclusions were made exclusively by the authors.

**References**

[1] Airbus, "Global Market Forecast 2025," Cirium, Jun. 2025. Accessed: Jul. 05, 2025. [Online]. Available: https://www.airbus.com/sites/g/files/jlcbta136/files/2025-06/Presentation%20GMF%202025-2044.pdf

[2] Boeing, "Commercial Market Outlook 2025-2044," Online, Jan. 2025. Accessed: Aug. 10, 2025. [Online]. Available: https://www.boeing.com/content/dam/boeing/boeingdotcom/market/assets/downloads/2025-commercial-market-outlook.pdf

[3] "FAST Flight Airworthiness Support Technology, Structural Blind Fasteners," *Airbus Technical Magazine*, vol. 64, Oct. 2019. Accessed: Jun. 10, 2025. [Online]. Available: https://aircraft.airbus.com/sites/g/files/jlcbta126/files/2021-11/Airbus-FAST64-magazine.pdf

[4] Collins Aerospace, "Thermoplastic Composites," Thermoplastic Composites. Accessed: Jul. 15, 2025. [Online]. Available: https://www.collinsaerospace.com/what-we-do/industries/commercial-aviation/aerostructures/advanced-structural-materials/thermoplastic-composites

[5] B. Parveez, M. I. Kittur, I. A. Badruddin, S. Kamangar, M. Hussien, and M. A. Umarfarooq, "Scientific Advancements in Composite Materials for Aircraft Applications: A Review," *Polymers*, vol. 14, no. 22, p. 5007, Nov. 2022, doi: 10.3390/polym14225007.

[6] F. Ozturk, M. Cobanoglu, and R. E. Ece, "Recent advancements in thermoplastic composite materials in aerospace industry," *J. Thermoplast. Compos. Mater.*, vol. 37, no. 9, pp. 3084–3116, Sep. 2024, doi: 10.1177/08927057231222820.

[7] B. A. Alshammari *et al.*, "Comprehensive Review of the Properties and Modifications of Carbon Fiber-Reinforced Thermoplastic Composites," *Polymers*, vol. 13, no. 15, p. 2474, Jul. 2021, doi: 10.3390/polym13152474.

[8] A. Siddique, Z. Iqbal, Y. Nawab, and K. Shaker, "A review of joining techniques for thermoplastic composite materials," *J. Thermoplast. Compos. Mater.*, vol. 36, no. 8, pp. 3417–3454, Aug. 2023, doi: 10.1177/08927057221096662.

[9] J.-H. Kweon, J.-W. Jung, T.-H. Kim, J.-H. Choi, and D.-H. Kim, "Failure of carbon composite-to-aluminum joints with combined mechanical fastening and adhesive bonding,"

*Compos. Struct.*, vol. 75, no. 1–4, pp. 192–198, Sep. 2006, doi: 10.1016/j.compstruct.2006.04.013.

[10] L. Larsen, M. Endrass, S. Jarka, S. Bauer, and M. Janek, "Exploring ultrasonic and resistance welding for thermoplastic composite structures: Process development and application potential," *Compos. Part B Eng.*, vol. 289, p. 111927, Jan. 2025, doi: 10.1016/j.compositesb.2024.111927.

[11] B. Mei and W. Zhu, "Accurate positioning of a drilling and riveting cell for aircraft assembly," *Robot. Comput.-Integr. Manuf.*, vol. 69, p. 102112, Jun. 2021, doi: 10.1016/j.rcim.2020.102112.

[12] O. S. Amosov, S. G. Amosova, and I. O. Iochkov, "Deep Neural Network Recognition of Rivet Joint Defects in Aircraft Products," *Sensors*, vol. 22, no. 9, p. 3417, Apr. 2022, doi: 10.3390/s22093417.

[13] I. F. Villegas, "Ultrasonic Welding of Thermoplastic Composites," *Front. Mater.*, vol. 6, no. 2019, 2019, doi: 10.3389/fmats.2019.00291.

[14] I. F. Villegas, "Strength development versus process data in ultrasonic welding of thermoplastic composites with flat energy directors and its application to the definition of optimum processing parameters," *Compos. Part Appl. Sci. Manuf.*, vol. 65, pp. 27–37, Oct. 2014, doi: 10.1016/j.compositesa.2014.05.019.

[15] S. K. Bhudolia, G. Gohel, K. F. Leong, and A. Islam, "Advances in Ultrasonic Welding of Thermoplastic Composites: A Review," *Materials*, vol. 13, no. 6, p. 1284, Mar. 2020, doi: 10.3390/ma13061284.

[16] I. F. Villegas and H. E. N. Bersee, "Ultrasonic welding of advanced thermoplastic composites: An investigation on energy-directing surfaces," *Adv. Polym. Technol.*, vol. 29, no. 2, pp. 112–121, Jun. 2010, doi: 10.1002/adv.20178.

[17] K. Wang *et al.*, "Characterization of weld attributes in ultrasonic welding of short carbon fiber reinforced thermoplastic composites," *J. Manuf. Process.*, vol. 29, pp. 124–132, Oct. 2017, doi: 10.1016/j.jmapro.2017.07.024.

[18] Y. Li, Z. Liu, J. Shen, T. H. Lee, M. Banu, and S. J. Hu, "Weld Quality Prediction in Ultrasonic Welding of Carbon Fiber Composite Based on an Ultrasonic Wave Transmission Model," *J. Manuf. Sci. Eng.*, vol. 141, no. 8, Aug. 2019, doi: 10.1115/1.4043900.

[19] Dominik Görick & Jonas Welsch, C. D. Gerardo, M. Kupke, E. Cretu, and R. Rohling, "Inline Monitoring of continous Ultrasonic Welding Processes of Thermoplastic Composites via a custom polyCMUT based Ultrasound Array," *E-J. Nondestruct. Test.*, vol. 29, no. 6, Jun. 2024, doi: 10.58286/29933.

[20] C. Kim, Y.-D. Shim, J. Kim, J. Gu, and E.-H. Lee, "Inspection of welded quality in thermoplastic welding using ultrasonics under different temperature conditions," *J. Mech. Sci. Technol.*, vol. 38, no. 10, pp. 5209–5218, Oct. 2024, doi: 10.1007/s12206-024-2304-1.

[21] M. Ju *et al.*, "Piezoelectric Materials and Sensors for Structural Health Monitoring: Fundamental Aspects, Current Status, and Future Perspectives," *Sensors*, vol. 23, no. 1, p. 543, Jan. 2023, doi: 10.3390/s23010543.

[22] A. S. Ergun, G. G. Yaralioglu, and B. T. Khuri-Yakub, "Capacitive Micromachined Ultrasonic Transducers: Theory and Technology," *J. Aerosp. Eng.*, vol. 16, no. 2, pp. 76–84, Apr. 2003, doi: 10.1061/(ASCE)0893-1321(2003)16:2(76).

[23] J. Joseph, B. Ma, and B. T. Khuri-Yakub, "Applications of Capacitive Micromachined Ultrasonic Transducers: A Comprehensive Review," *IEEE Trans. Ultrason. Ferroelectr. Freq. Control*, vol. 69, no. 2, pp. 456–467, Feb. 2022, doi: 10.1109/TUFFC.2021.3112917.

[24] C. D. Gerardo, E. Cretu, and R. Rohling, “Fabrication and testing of polymer-based capacitive micromachined ultrasound transducers for medical imaging,” *Microsyst. Nanoeng.*, vol. 4, no. 1, p. 19, Dec. 2018, doi: 10.1038/s41378-018-0022-5.
[25] J. Welsch, E. Cretu, R. Rohling, and C. D. Gerardo, “Ultrathin, High Sensitivity Polymer-based Capacitive Micromachined Ultrasound Transducers (polyCMUTs) for Acoustic Emission Sensing in Fiber Reinforced Polymers,” in *2022 IEEE International Ultrasonics Symposium (IUS)*, Venice, Italy: IEEE, Oct. 2022, pp. 1–5. doi: 10.1109/IUS54386.2022.9958226.
[26] S. Frey, S. Vostrikov, L. Benini, and A. Cossettini, “WULPUS: a Wearable Ultra Low-Power Ultrasound probe for multi-day monitoring of carotid artery and muscle activity,” in *2022 IEEE International Ultrasonics Symposium (IUS)*, Venice, Italy: IEEE, Oct. 2022, pp. 1–4. doi: 10.1109/ius54386.2022.9958156.
[27] M. Angerer, J. Welsch, C. D. Gerardo, E. Cretu, and R. Rohling, “Accurately Predicting the Performance of Polymer-Based CMUTs by Coupling Finite-Element and Analytical Models,” *IEEE Open J. Ultrason. Ferroelectr. Freq. Control*, vol. 4, pp. 227–230, 2024, doi: 10.1109/ojuffc.2025.3526123.
[28] M. Angerer, J. Welsch, C. D. Gerardo, N. V. Ruiter, E. Cretu, and R. Rohling, “Studying the Effects of Mutual Acoustic Impedance on the Performance of Polymer-Based CMUTs,” *IEEE Open J. Ultrason. Ferroelectr. Freq. Control*, Forthcoming 2024, doi: 10.1109/OJIM.2022.1234567.
[29] M. U. Khan, M. La Mura, M. Saccher, R. Van Schaijk, R. Dekker, and A. S. Savoia, “Fast and Accurate Estimation of Collapse and Snapback Voltages of CMUTs,” in *2024 IEEE Ultrasonics, Ferroelectrics, and Frequency Control Joint Symposium (UFFC-JS)*, Taipei, Taiwan: IEEE, Sep. 2024, pp. 1–4. doi: 10.1109/UFFC-JS60046.2024.10793933.
[30] J. Koyanagi, M. Takamura, K. Wakayama, K. Uehara, and S. Takeda, “Numerical simulation of ultrasonic welding for CFRP using energy director,” *Adv. Compos. Mater.*, vol. 31, no. 4, pp. 428–441, Jul. 2022, doi: 10.1080/09243046.2022.2031376.
[31] F. Teston, C. Meynier, E. Jeanne, N. Felix, and D. Certon, “P2P-8 Characterization Standard of CMUT Devices Based on Electrical Impedance Measurements,” in *2006 IEEE Ultrasonics Symposium*, Vancouver, BC, Canada: IEEE, 2006, pp. 1963–1966. doi: 10.1109/ultsym.2006.496.
[32] A. M.-E. Dorado, “Composite Material Characterization using Acoustic Wave Speed Measurements,” presented at the American Society for Non-Destructive Testing Annual Conference, Salt Lake City, Utah, USA: American Society for Non-Destructive Testing, 2015. [Online]. Available: https://www.osti.gov/servlets/purl/1326353
[33] *MIL-HDBK-1823A: Nondestructive Evaluation System Reliability Assessment. Department of Defense Handbook.*, Washington, DC, USA., 2009.